\documentclass[letterpaper]{article}
\usepackage{spconf,amsmath,graphicx,url}
\usepackage{xcolor}
\usepackage{amssymb}

\title{Tensor-based grading: a novel patch-based grading approach for the analysis of deformation fields in Huntington's disease}
\name{Kilian Hett$^1$, Hans Johnson$^2$, Pierrick Coup\'{e}$^3$, Jane S. Paulsen$^{4,5}$, Jeffrey D. Long$^{5,6}$, and Ipek Oguz$^1$\thanks{This work was supported, in part, by the NIH grant R01-NS094456. The  PREDICT-HD study was funded by the National Center for Advancing Translational Sciences, and the National Institutes of Health (NIH; NS040068, NS105509, NS103475) and CHDI.org. Vanderbilt University Institutional Review Board has approved this study.}}
\address{$^1$Vanderbilt University, Dept. of Electrical Engineering and Computer Science, Nashville TN, USA \\
$^2$ University of Iowa, Dept. of Electrical and Computer Engineering, Iowa City, IA, USA\\
$^3$ CNRS, University of Bordeaux, Bordeaux INP, LABRI, UMR5800, PICTURA, France\\
$^4$ University of Iowa, Dept. of Neuroscience,  Iowa City IA, USA\\
$^5$ University of Iowa, Dept. of Psychiatry, Iowa City IA, USA\\
$^6$ University of Iowa, Dept. of Biostatitsics, Iowa City IA, USA\\
}
\begin{document}
%\ninept
%
\maketitle
\begin{abstract}
% \xxx{did we check whether this needs to be anonymous?}
The improvements in magnetic resonance imaging have led to the development of numerous techniques to better detect structural alterations caused by neurodegenerative diseases. Among these, the patch-based grading framework has been proposed to model local patterns of anatomical changes. This approach is attractive because of its low computational cost and its competitive performance. Other studies have proposed to analyze the deformations of brain structures using tensor-based morphometry, which is a highly interpretable approach. In this work, we propose to combine the advantages of these two approaches by extending the patch-based grading framework with a new tensor-based grading method that enables us to model patterns of local deformation using a log-Euclidean metric. We evaluate our new method in a study of the putamen for the classification of patients with pre-manifest Huntington's disease and healthy controls. Our experiments show a substantial increase in classification accuracy ($87.5\pm 0.5$ vs.~$81.3\pm0.6$) compared to the existing patch-based grading methods, and a good complement to putamen volume, which is a primary imaging-based marker for the study of Huntington's disease. 
% \xxx{not a word - maybe `a good complement to putamen volume'?}
% with putamen volume 
% \xxx{it feels odd that your first keyword doesn't appear in the abstract}
% Max 250 words
% However, this exemplar-based approach can also be impacted by the quality of training dataset. Indeed, exemplar-based methods see reduced detection performance when evaluated using highly heterogeneous dataset. 
\end{abstract}

\section{Introduction}
% \xxx{now i don't think examplar appears anywhere outside of your abstract :) do you feel strongly about using this term?}
% \xxx{I worry about calling it detection (in the abstract, and I imagine in the rest of the text), because there's already a method that can detect HD with 100\% accuracy. Maybe classification of MRIs from premanifest HD patients vs matching healthy controls?}
% Huntington disease --> Good disease candidate for the development of new imaging-based technique (perfect ground truth - diagnosis)
Huntington's disease (HD) is a fatal autosomal dominant inherited disorder that causes motor, behavioral and cognitive symptoms.
% leading to death in around 15 years after the first clinical onset. 
% The pathological mutation consists of an abnormal cytosine, adenine, and guanine (CAG) repeat in the huntingtin gene.
% \cite{dayalu2015huntington}. 
Imaging studies have shown structural changes in the striatum \cite{ross2014huntington}. Moreover, it has been demonstrated that the volume of the putamen is a sensitive imaging-based marker for the tracking of changes that occur in the lifespan of patients with HD \cite{paulsen2014clinical}.
% \xxx{can be removed if you need space} 
Unlike many other degenerative diseases,  gene status can be established via a genetic test
% the diagnosis can be made with genetic testing 
well before the onset of first symptoms, which makes HD a good candidate for the evaluation of new imaging-based methods. 
% \xxx{should cite something from the sarah tabrizi group too}
% Medical imaging --> New methods to detect early

Numerous advanced methods have been developed to detect the structural modifications earlier in the course of neurodegenerative diseases using anatomical MRI \cite{arbabshirani2017single}.
On the one hand, studies have proposed to perform analysis based on MRI intensity. Among these methods, the patch-based grading framework has shown attractive characteristics such as low computational cost and competitive performance to recent methods based on deep-learning \cite{coupe2012scoring,tong2017novel}. 
In addition, patch-based grading methods also provides the localization of structural differences.
% \xxx{interpretability}
%  \yyy{you later talk about how your method is interpretable, unlike traditional patch-based; so this feels a bit  contradictory}
Indeed, these approaches aim to measure the local similarity of anatomical patterns by comparing the MRI under study to a template library composed of MRIs representing two distinct populations. Patch-based grading has been successfully applied to various intensity-based features, such as gray matter density maps \cite{komlagan2014anatomically}, features extracted using texture filters \cite{hett2018adaptive}, and diffusion-weighted imaging parameters \cite{hett2019multimodal}.
% Advanced patch-based grading  --> Deformation field
On the other hand, many studies have proposed to assess anatomical changes among MRIs with the analysis of the deformation tensor fields resulting from image registration \cite{ashburner2000voxel}. Features derived from deformation tensors are highly interpretable since they provide both localization and geometric characteristic of the brain alterations. Indeed, these methods register the MRIs under study into the same stereotaxic space and use the resulting deformation fields to capture regional changes of brain structures (\emph{e.g.},  local expansion or shrinkage). In particular, the log-Euclidean approach has been proposed as an efficient Riemannian framework for tensor-based morphometry \cite{arsigny2006log}. The efficiency of tensor-based approaches has been demonstrated in diverse applications such as disease detection, longitudinal studies, clinical trials, and stability control of acquisition protocols \cite{hua2011accurate,leow2006longitudinal,koikkalainen2011multi}. 
% \xxx{since you have the space in ref page, add a ref for each application you list in this sentence}
% \yyy{you keep saying it is efficient but I think you need one sentence crisply describing what's good about tensor-based morphometry. you said twice that patch-based is good for interpretability, low-cost, good performance. tensor-based is good because ...?}

% \xxx{i'm changing all of the "works" to "studies" - "works" feels very awkward to me. might want to check with a native speaker}
% Contribution
% \xxx{not sure how this follows}
In this paper, we propose to combine the respective advantages of patch-based grading and tensor-based morphometry in a novel tensor-based grading framework. We evaluate our method with the classification of pre-manifest HD patients and control subjects. The performance of our new method is compared with the volume of the putamen, the original patch-based grading using T1w intensities, and a recent texture-based grading approach. In addition, we also investigate the complementary nature of these different methods. These experiments show that our new method has competitive performance and also suggest complementarity with volumetric features.

\section{Materials and Methods}
\subsection{Dataset}
\label{sec:data}
All T1-weighted (T1w) MRIs come from the PREDICT-HD study \cite{paulsen2008detection}, which is a multi-site longitudinal study of HD. The MRIs have been acquired using 3 Tesla MRI scanners from different vendors (\emph{e.g.}, GE, Phillips, and Siemens).
% using inversion recovery TurboFLASH sequence.
% (MPRAGE)
% The main parameters of the MR acquisition sequences included in this study vary as follows: TR: 8$–-$8.4sec, TE: 3.5$-$3.8sec, TI: 826$–-$843sec, matrix: 256$\times$256, voxel size: 1$\times$1mm, slice thickness: 0.9$–-$1.2mm. 
The cohort used in the study includes 683 MPRAGE images from subjects representing three populations: control subjects (CN), pre-manifest HD that is composed of subjects with the expanded cytosine-adenine-guanine (CAG) repeat but who have not yet had a motor diagnosis at the time of the scan, and manifest HD which refers to patients who already have a motor diagnosis by the time of the scan (see Table~\ref{tab:dataset}).
% an abnormal CAG repeat but who have not developed clinical onset at the scan, and manifest HD which refers to patients who suffer from clinical onset at the scan (see Table~\ref{tab:dataset}).
% This dataset is composed of HD patients having cytosine, adenine, and guanine (CAG) from 41 to 43 repeats. 
Only subjects with CAG length from 41 to 43 repeats and at least 2 longitudinal scans at 3T have been embedded in this study. 
% \xxx{you need to mention you're selecting 41-43 only}
% \xxx{I thought I already said this, but: you need to describe how you chose the 683 out of the much larger PREDICT-HD: the subjects that have 2+ longitudinal scans acquired at 3T. Also, you're restricting to 41-43 CAGs.}
% provides a description of the dataset used in our experiments. 
\begin{table}[t]
\centering
\caption{Demographic description of the PREDICT-HD dataset used in our experiments.}
\label{tab:dataset}
\begin{tabular}{l @{\hspace{0.1cm}}|@{\hspace{0.1cm}} c @{\hspace{0.1cm}}|@{\hspace{0.1cm}} c @{\hspace{0.1cm}}|@{\hspace{0.1cm}} c }
					  &  Control			& \multicolumn{2}{c}{Huntington's disease} \\ \cline{3-4}
				      &		 	 		    & Pre-manifest 		& Manifest  			\\ \hline
Number of MRIs 		  & 327 	     		&  300	         	& 56  				\\
Age (years) 	   	  & 	49.3$\pm$11.9 	&  43.4$\pm$10.1		& 56.8$\pm$6.5 		\\
Sex (F/M)   	      & 	206/121     		&  199/101	  		& 28/28	 			\\
CAG length  	      & 	15-35    		&  41-43 			& 41-43				\\
\end{tabular}
\end{table}

\subsection{Preprocessing}
The preprocessing has been conducted with the BRAINSAutoWorkup pipeline \cite{pierson2011fully}.
% \xxx{i am not sure about this paper, i think it should be https://www.ncbi.nlm.nih.gov/pmc/articles/PMC3827877/ i actually think there is a more recent autoworkup apper with regina kim as first author (but not a workshop paper, a journal paper) but i cant seem to find it right now}
This pipeline is composed of the following steps: (1) denoising with non-local means filter, (2) anterior/posterior commissure and intra-subject alignments with rigid transformation, (3) bias-field correction, and (4) regional segmentation mask with a multi-atlas method using atlases from Neuromorphometrics\footnote{\url{http://www.neuromorphometrics.com}}. 
After rigid alignment, we take the union of the putamen masks of all subjects. The bounding box of this union is used as a region of interest (ROI). All images are cropped with this common ROI.
%The union of putamen segmentation mask over the entire dataset after rigid alignment has been used to crop the images around the putamen of each hemisphere.  
% \xxx{do you mean left/right? not clear what you're union-ing}
% \xxx{i really strongly recommend grammarly - i'm seeing a lot of errors it would have caught}

\subsection{Deformation-based tensor computation}\label{sec:deformation}
A non-rigid symmetric normalization registration \cite{avants2011reproducible} has been conducted to estimate deformation fields in the ICBM 152 nonlinear stereotaxic space\footnote{\url{http://www.bic.mni.mcgill.ca/ServicesAtlases/ICBM152NLin2009}} .
% . After the deformation fields have been obtained from a non-rigid registration using the 
% \xxx{weird sentence - "after X, Y", but there's no "Y". do you mean "next, X"?}. 
% \xxx{since this whole thing rides on the quality of your registration, you have to say a few words about how you registered the images. Ants? what parameters? what metric? etc} 
Next, the Jacobian matrix $J$ of the resulting deformation field, defined at each voxel as:
\begin{equation}
	J = \begin{pmatrix} 
			\frac{\partial x-u_x}{\partial x} & \frac{\partial x-u_x}{\partial y} & \frac{\partial x-u_x}{\partial z}\\
			\frac{\partial y-u_y}{\partial x} & \frac{\partial y-u_y}{\partial y} & \frac{\partial y-u_y}{\partial z}\\	
			\frac{\partial z-u_z}{\partial x} & \frac{\partial z-u_z}{\partial y} & \frac{\partial z-u_z}{\partial z}\\	
		\end{pmatrix},
\end{equation}
has been used to compute the deformation-based tensor $\Phi = \sqrt{J^{T}J}$ that describes the geometry of local deformation at each voxel.
% \xxx{can prob remove the equation 1 if you are in dire need of space}

\begin{figure}[t]
\includegraphics[width=1\linewidth]{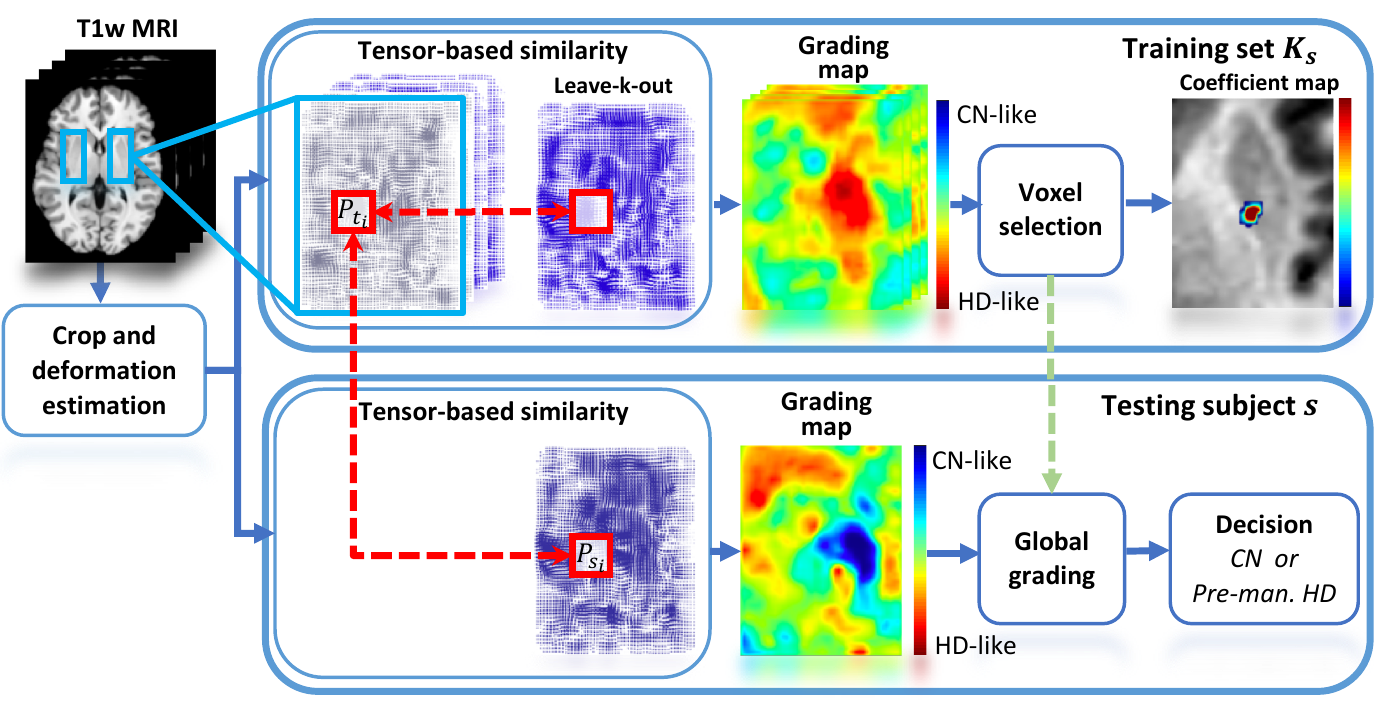}	
% \xxx{i don't think you refer to this figure anywhere} Description of the pipeline used to compute the proposed tensor-based feature. 
\caption{On the top row: a template library $K_s$, composed of CN (gray color) and manifest HD (blue color), is built for each subject $s$ using an age matching preselection. For each template $t$ of $K_s$ the tensor-based grading map is estimated following a leave-k-out cross-validation procedure. Once the grading maps are estimated, a sparse regularization model is computed to estimate the voxels having the most discriminant grading values. On the bottom row: for each subject $s$ the proposed tensor-based grading method is computed using the template library $K_s$. Then the most discriminant voxels based on $K_s$ are used to compute a global grading feature which is used to make the final decision.}\label{fig:pipeline}
% The tensor-fields First, the deformation-based tensor fields are estimated by a non-rigid registration of T1w MRI (Sec.~\ref{sec:deformation}). Then, the estimated tensor maps are used to compute the tensor-based similarity at each voxel (Sec.~\ref{sec:grading}). A leave-k-out cross-validation procedure is used to compute patch-based grading for the MRIs that compose the template library. A sparse regularization model is used to select and aggregate the voxels with the most discriminant grading values into a global grading feature, which is used to make the classification decision (Sec.~\ref{sec:global}).}
% \xxx{updates we talked about}}
% \xxx{let's talk about this caption}
% \xxx{i would keep the N's in the figure - i don't think you explicitly say how many in test set anywhere else}}
% \xxx{it might be nice to put (sec 2.x) at the end of each sentence in this caption}
\end{figure}

\subsection{Tensor-based grading}\label{sec:grading}
Once the deformation-based tensor field is computed, local similarities are estimated at each voxel to estimate the degree of the structural changes (see Fig.~\ref{fig:pipeline}). We use a patch-based grading approach \cite{coupe2012scoring} instead of analyzing each tensor independently of its neighborhood, in contrast to tensor-based morphometry \cite{ashburner2000voxel,arsigny2006log}. We hypothesize that such a patch-wise analysis will enable us to better model the deformation patterns. 

First, for each subject $s$, a template library $K_s$ composed of the tensor fields $t$ from CN and manifest HD patients is built using an age matching preselection. This preselection technique aims to reduce the bias introduced by age-related differences.
% \xxx{how? randomly? any constraints?}. 
% \xxx{i don't think you explain $K_{x_i}$ - this reads like there's a single library K. also, i would use x as spatial location rather than subject, to avoid confusion} 
% \yyy{there are still some problems with the notation. you never say what $t_i$ means, and in 3 places you say $t_i \in K_s$ or $t_i \in K_s_i$ when i think you mean $t \in K_s$}
Then, at each voxel $i$ (\emph{i.e.}, image coordinate) of the subject $s$, the patch-based grading is computed as,
\begin{equation}
g_{s_i} = \frac{\sum_{t \in K_{s}} \exp \Big( - \tfrac{d(P_{s_{i}}, P_{t_{i}})}{h_i}  \Big) y_t}{\sum_{t \in K_{s}} \exp \Big( - \tfrac{d(P_{s_{i}}, P_{t_{i}})}{h_i}  \Big)}, \label{eq:grade}
\end{equation} 
where $h_i = min(d(P_{s_{i}},P_{t_{i}})), \forall t \in K_{s}$ is used to normalize the patch similarity at each voxel. $y_t$ is a binary indicator of the pathological status of the template $t$, set to $-1$ for manifest HD and $+1$ for control. The distance $d(P_{s_{i}},P_{t_{i}})$ is based on the log-Euclidean framework \cite{arsigny2006log}. $d$ describes the similarity of two tensor patches $P_{s_{i}}$ and $P_{t_{i}}$ surrounding voxel $i$ and is defined as follows:
% \xxx{is your definition of d same as d'arsigny? if so, reference here.}:
\begin{equation}
d(P_{s_{i}},P_{t_{i}}) = \sum_{n=1}^{N} Trace ( (log \Phi_{{s_i}}(n) - log \Phi_{{t_i}}(n))^2 ) ^{\frac{1}{2}},\label{eq:weight}
\end{equation}
%where $N$ is the number of voxels in each patch, and $n$ is the location (\emph{i.e.}, patch coordinate) within the patch of tensors, such that $\Phi_{{s_i}}(n)$ and $\Phi_{{t_i}}(n)$ are individual tensors in the patches $P_{s_{i}}$ and $P_{t_{i}}$ in subject $s$ and template $t$, respectively.
where $N$ is the number of voxels in each patch, and $n$ is the index (ranging from $1..N$) within the patch of tensors centered around voxel $i$, such that $\Phi_{{s_i}}(n)$ and $\Phi_{{t_i}}(n)$ are individual tensors in the patches $P_{s_{i}}$ and $P_{t_{i}}$ in subject $s$ and template $t$, respectively.
% \xxx{let's talk about this notation in person, i have a lot of questions}

\subsection{Feature selection and global grading computation}\label{sec:global}
Since the images have been deformably registered into the same space, we can compare the grading values at a given voxel across the set of images and use these as features for classification. This further allows us to use feature selection techniques to select the most discriminant voxels \cite{tong2017novel}.
%The accurate alignment of each voxel enables the use of feature selection techniques to select the most discriminant voxel using the grading values computed previously.
% \xxx{not sure what you mean in the last sentence - what does registration accuracy have to do with feature selection?} 
In our work, we used an elastic-net regularization model that provides a sparse representation of the most discriminative features, defined as follows:
\begin{equation}
 \hat{\beta} = \underset{\beta}{argmin} \frac{1}{2} || G\beta - Y ||^2_2 + \rho ||\beta||^2_2 + \lambda ||\beta||_1, \label{eq:en}
\end{equation}
where $\hat{\beta}$ represents the regularization coefficients computed using the tensor-based grading maps $G$ composed of the $g_{t_i}$ as defined in Eq.~\ref{eq:grade}, computed on an inner fold of the cross-validation within the template library $K_s$. $Y$ represents the vector of pathological status composed of the binary indicators $y_t$ defined in Sec.~\ref{sec:grading}.
% and $G$ is the matrix of grading values for the template library $K_s$. 
% \xxx{I don't understand the first half of the sentence - isn't G `the tensor-based grading map from the trainign templates $K_s$'? why aren't we using $G$ instead of repeating that part twice in one sentence?}
% Each row of $G$ represents the grading values of the templates library and columns represent the tensor-grading values for each voxel. 
Finally, a global tensor-based grading feature for the subject $s$ is computed as,
\begin{equation}
	g_s = \frac{\sum_{i \in s}\hat{\beta}_i g_{s_i}}{\sum_{i \in s}\hat{\beta}_i}.
\end{equation}
Thus, $g_s$ represents a measure that estimates the global deformation differences of the subject $s$ under study. 
% \xxx{This has become super confusing, i liked the older version much better. What does the set of localization mean? Do you mean $\Omega$ is the image domain? If you are going to bring in the image domain, shoudln't we do it much much (much) earlier when you first define patches and whatnot, rather than at the very end? i don't get it.}\yyy{why does it have to be an abnormality? If your template library consists of, e.g. male and female, or PD and ET, there's no normal/abnormal to speak of. so i don't think you can call it an abnormality when you talk about the method generally (it's fine in the results, since this particular study does have normals vs abnormals). i fixed it elsewehere in the text but not sure how to fix this instance.}

\subsection{Evaluation and implementation details}
We evaluate our new method by classifying pre-manifest HD patients and control subjects. For each subject $s$, the template library $K_s$ has been built using 100 deformation-based tensor fields, 50 from CN MRI and 50 from manifest HD MRI %using an age matching pre-selection.
pre-selected via age matching to the subject $s$.
% \xxx{The template library $K_s$ for each subject $s$ is a subset of these 100 tensor fields. or is the $K_s$ equal to the entire set of 100, except for same-subject issues you mention next?} 
Moreover, all longitudinal scans of a given subject $s$ have been removed from its template library $K_s$ to avoid double-dipping.
% \xxx{i edited the prev sentence - is this true?}
% \xxx{so you're only using one scan per patient, period? or you're just not allowing the later scans of the patient currently under study? K or $K_x$, i guess, is what I'm asking}
The tensor fields have been estimated using a non-rigid diffeomorphic registration computed using ANTs \cite{avants2011reproducible}. The registration has been performed using the cross-correlation metric and the smoothing parameter has been set to 2mm.
The elastic-net regularization model has been computed using the Matlab Statistic and Machine Learning toolbox, with $\rho = 0.2$ and $\lambda = 0.09$ (see Eq.~\ref{eq:en}). The classification has been computed using a linear support vector machine with the soft margin parameter set to $C=1$. Finally, a stratified cross-validation procedure iterated 100 times has been conducted.

% (see Table~\ref{tab:size}, \ref{tab:comparison}, and \ref{tab:combination}) \xxx{i woudl remove - it's too early to talk about results!}.
% \xxx{the results section looks like a single very long paragraph, i'll put some breaks but see if they make sense to you}

\subsection{Comparison with state-of-the-art}
% \xxx{heavily edited paragraph, please check}
To evaluate the performance of the proposed tensor-based grading method, we conduct single-feature and two-feature classification experiments. The features compared are the volume of putamen using the multi-atlas label fusion segmentation mask obtained from the BRAINSAutoWorkup pipeline \cite{pierson2011fully}, patch-based grading using T1w intensity \cite{coupe2012scoring}, and a patch-based grading approach that fuses grading maps estimated from different texture maps \cite{hett2018adaptive}. These latter two features have been aggregated within the putamen segmentation mask to obtain the global grading. Moreover, for these two patch-based grading methods, a piece-wise linear histogram standardization has been used to normalize intensities.
% \xxx{not clear who 'both' refers to} 
% and the proposed patch-based grading using tensors. 
% \xxx{a sentence about intensity normalization for intensity and texture-based} 
The mean accuracy (ACC), sensibility (SEN), and specificity (SPE) over the 100 iterations are reported.

\section{Results and Discussion}
% \xxx{unless you mean different things by "deformation-based" vs "tensor-based grading", choose one, stick to it}
% To evaluate our new tensor-based grading framework, we conducted three experiments. We investigated the impact of different patch sizes. Then, our new tensor-based grading has been compared to putamen volume and two traditional patch-based grading methods. Finally, the combination of different features has been compared to investigate possible complementarity of these different methods. 
% Finally, in addition, the putamen areas impacted by the most discriminant deformation are presented.

% \xxx{is there a reason you want the table placements to be [!ht]? it's very difficult to separate caption from text, esp since they're all same font size. i would move the tables to [t] so the captions are always separated from text. otherwise maybe make captions \footnotesize{small} so they look visually different?}

\begin{table}[b]
\centering
\caption{Patch size influence on the classification performance. All results are expressed in percentage.}\label{tab:size}
\begin{tabular}{l|c|c|c}
	  		  & Accuracy  	  		  & Sensitivity 			  & Specificity   \\ \hline 
1x1x1 (voxel) & 82.1$\pm$0.5  		  & 81.7$\pm$0.7    		  & 82.5$\pm$0.8  \\
3x3x3 (patch) & \textbf{87.5$\pm$0.5} & \textbf{88.2$\pm$0.7}     & \textbf{86.9$\pm$0.6} \\ 
5x5x5 (patch) & 86.1$\pm$0.5	  	  & 86.0$\pm$0.7  	 		  &	86.2$\pm$0.7  		\\
\end{tabular}
\end{table}

% \xxx{consider changing each of the first sentences to a header: {\bf\\ul Experiment 1: Effect of patch size on tensor-based grading.} Experiment 2: single-feature classification. Experiment 3: two-feature classification. i don't know why \\ul is not working for underlining, prob missing a package...}

% \xxx{i think it's percentage points in english, not points of percentage}
First, we investigated the effect of patch size for the classification of pre-manifest HD patients and control subjects. Table~\ref{tab:size} summarizes the obtained results. We compared voxel-wise (\emph{i.e.}, 1$\times$1$\times$1 voxel) and patch-wise (\emph{i.e.}, 3$\times$3$\times$3 and 5$\times$5$\times$5 voxels) tensor-based grading. The results of this experiment demonstrate the importance of embedding neighbor deformations. Indeed, patch-wise tensor-based grading using patches of 3$\times$3$\times$3 voxels obtained in average 87.5\% accuracy, which improves the tensor-based grading using single voxel-wise similarity by 5 percentage points in accuracy and specificity, and 7 percentage points in  sensitivity. In our experiments, we limited our investigation to patch sizes up to 5$\times$5$\times$5 voxels since the computational cost becomes prohibitive. 
% for larger patches.
% \xxx{this will raise question about what the computational cost is - maybe squeeze in a sentence about that somewhere}

% \xxx{i vaguely want to switch the order of tables 2 and 3 (and the paragraphs describing them. table 3 is your main selling point. it's harder to get excited about parameter space exploration (table 2) before we know this is worth doing (table 3). just a thought.} 

\begin{table}[t]
\centering
\caption{Comparison of different patch-based grading methods. The volume of putamen is given as baseline. All results are expressed in percentage.}\label{tab:comparison}
\begin{tabular}{l@{\hspace{0.1cm}}|@{\hspace{0.1cm}}c@{\hspace{0.1cm}}|@{\hspace{0.1cm}}c@{\hspace{0.1cm}}|@{\hspace{0.1cm}}c}
			            & Accuracy 			    & Sensitivity 	        & Specificity 			\\ \hline
Putamen volume          & 82.9$\pm$0.5 		    & 84.3$\pm$0.6          & 81.3$\pm$0.6 			\\
Intensity-based grading & 73.8$\pm$0.7          & 77.0$\pm$1.0          & 70.8$\pm$0.7 \\
Texture-based grading   & 81.3$\pm$0.6          & 82.1$\pm$0.7          & 80.7$\pm$0.9 \\
Proposed method	        & \textbf{87.5$\pm$0.5} & \textbf{88.2$\pm$0.7} &   \textbf{86.9$\pm$0.6} 
\end{tabular}
\end{table}

% \xxx{Maybe you should move the compared methods to a subsection 2.x? the current placement feels too late in the paper}
Second, the proposed tensor-based grading method was compared to putamen volume, which is provided as baseline, the original patch-based grading using T1w intensities \cite{coupe2012scoring}, and a texture-based grading approach \cite{hett2018adaptive}. The results are summarized in Table~\ref{tab:comparison}. 
The classification performance of intensity-based grading is low compared to the volume of putamen. 
% \xxx{the part i highlighted in red - i would stick to the name you use in the table, or change the name you use in the table, it gets hard to follow otherwise}.
This result is unexpected given the detection performance of this approach obtained for other neurodegenerative diseases (\emph{e.g.}, Alzheimer's disease \cite{coupe2012scoring}). We note that this low performance may be caused by the highly heterogeneous nature of the dataset used to evaluate our method. As described in Sec.~\ref{sec:data}, the PREDICT-HD dataset includes MRIs from diverse scanners and acquisition sequences \cite{paulsen2008detection}. The grading performance is somewhat improved by the use of texture, which obtains comparable accuracy to the volume of putamen.
% (\emph{i.e.}, 81.3\% of accuracy compared to 82.9\% for the volume \xxx{why mention this? you didn't mention the numbers for original grading. it's all in the table}). 
In contrast, our novel tensor-based grading obtains 87.5\% accuracy for the classification of pre-manifest HD patients. This improves the accuracy by 5 percentage points compared to putamen volume and texture-based grading and by 13 percentage points compared to intensity-based grading.
% \xxx{i heavily edited the next sentence, can you check if it still makes sense? :)}
We hypothesize that the superior performance of tensor-based grading compared to traditional grading approaches may partially come from the use of tensor fields instead of features derived from MRI intensity, since deformation-based tensor fields might be less impacted by the heterogeneous nature of acquisition sequences than intensity-based features. Tensor-based grading also provides a potentially more interpretable description of the structural differences since it indicates not only the localization but also the local geometry of the detected differences.
% \xxx{i don't know what you mean here. it can be better description of structure abnormality because it measures structure abnormality?}. 
Another advantage of this method is that it does not require an explicit segmentation of the structure. 
%consists in the independence of structure segmentation since our method does not need segmentation of the region of interest.

\begin{table}[b]
\caption{Comparison of two-feature classification results. All results are expressed in percentage.}\label{tab:combination}
\centering
\begin{tabular}{l|c|c|c}
			      & Accuracy 				& Sensitivity 	 &  Specificity 					\\ \hline
Volume + Texture  & 84.4$\pm$0.6 			& 84.6$\pm$0.7 	 &  84.3$\pm$0.9 					\\
Volume + Tensor   & \textbf{89.1$\pm$0.4} 	& \textbf{88.5$\pm$0.6}	 &  \textbf{89.6$\pm$0.7} 	\\
Tensor + Texture  & 87.9$\pm$0.4 			& 88.4$\pm$0.7 	 &  87.4$\pm$0.6 					\\
\end{tabular}
\end{table}

Next, as described in Table~\ref{tab:combination}, the complementarity of the different methods for the classification of pre-manifest HD patients and control subjects has been investigated. The results of this experiment show that the  combination of volume with texture-based grading, and volume with tensor-based grading improve the classification performance compared to the use of single feature. Indeed, the combination of putamen volume and tensor-based grading obtains 89.1\%  accuracy, which mainly benefits from a higher specificity than the tensor-based grading method alone. In contrast, the combination of texture and tensor-based features does not improve classification performance.

\begin{figure}[t]
\includegraphics[width=1\linewidth]{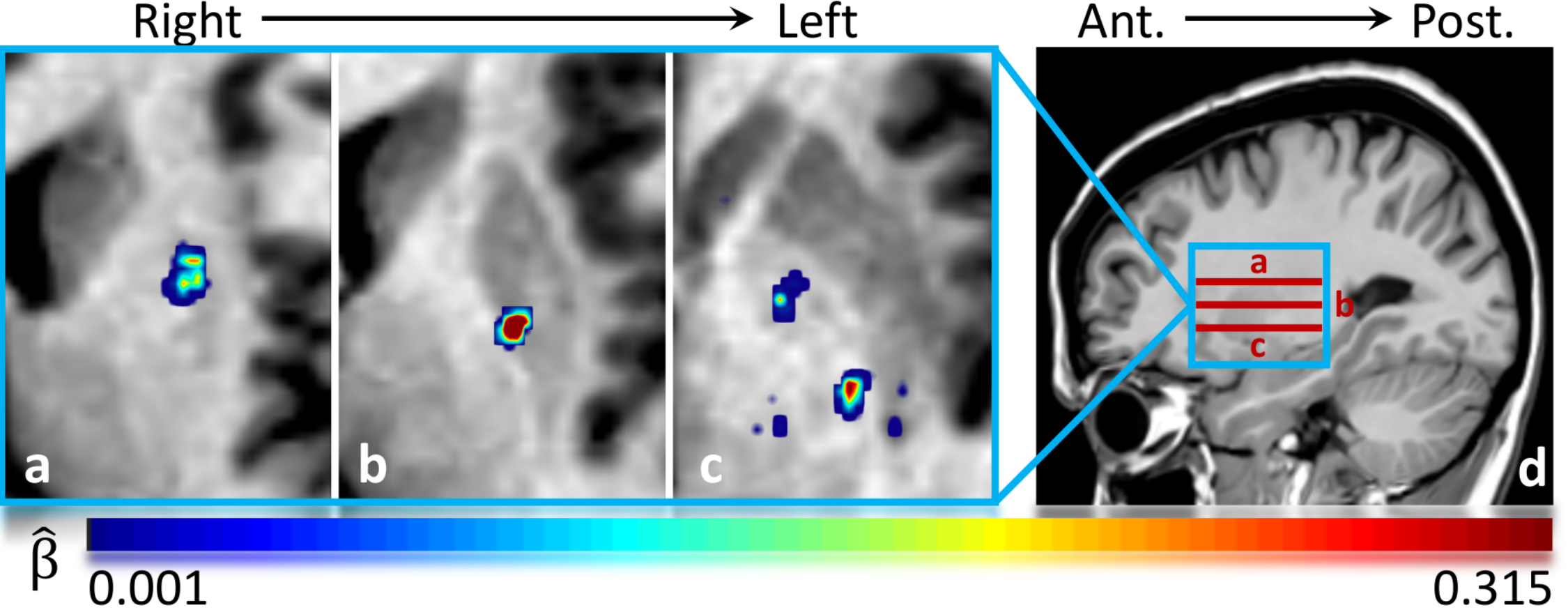}
\caption{{\bf (a-c)} Three axial slices showing $\hat{\beta}$ (see Eq.~\ref{eq:en}) as color overlay. Voxels with $\hat\beta=0$ are not shown for clarity. {\bf (d)} The localization of slices shown in (a-c).}
%From left to right, three slices using the coronal planes of the coefficient map computed in Eq.~\ref{eq:en} from the left putamen, and on the right side, the localization of each slices of coefficient using the sagittal plane. The color-bar is provided, in blue: voxels having a coefficient near zero, and in red: voxels having the highest coefficients.\xxx{let's also talk about this figure}}
\label{fig:coefmap}
\end{figure}

Finally, Fig.~\ref{fig:coefmap} presents the coefficient map of the left putamen area estimated from Eq.~\ref{eq:en}. This illustrates the interpretability of the results obtained from the proposed method. Indeed, our method enables us to highlight the areas where the most discriminant deformations occur.
% \xxx{That's not true, is it? if I understand what you're showing correctly, you're highlighting the most discriminating features, not the most severe deformations}. 
In our work, results indicate a discriminant deformation 
% \xxx{again, not sure that's true} 
of the superior part of putamen (Fig.~\ref{fig:coefmap}-a). Our results also show an abnormal deformation on the medial side of the putamen (Fig.~\ref{fig:coefmap}-b), and around the globus pallidus (Fig.~\ref{fig:coefmap}-c). These results are in line with previous imaging-based anatomical studies that used volumetric measurements \cite{younes2014regionally}. 
It is noteworthy that while we focused our current experiments on the putamen, nevertheless we also found discriminant features in the surrounding structures such as the globus pallidus. In future work, we will extend this work to whole brain analysis. 
%In future work, further investigations will be conducted to better analyze these sub-cortical structures. %\xxx{This is interesting. See what you think of the following argument, and if you decide to keep it, remove the last sentence from two paragraphs ago ("another advantage...", since I'm repeating it here).} \xxx{Another advantage of this method is that it does not require an explicit segmentation of the structure. Not only does this prevent measurement noise due segmentation inaccuracy, but also, perhaps more importantly, it allows for detection of discrimination features outside the initial region of interest (ROI), which is useful for exploration and hypothesis generation. While we focused on the putamen as the initial ROI for these experiments, we nevertheless were able to also identify discriminating features in the neighboring pallidus. This may partially contribute to the superior performance of our approach compared to the putamen volume only (Table \ref{tab:comparison}). }

\section{Conclusion}
In this work, we proposed a novel tensor-based grading method for the analysis of tensor fields obtained from non-rigid registration. The proposed method has been evaluated with the classification of pre-manifest HD and control subjects by analyzing the putamen area. Our approach has shown an increase of performance compared to previous patch-based approaches based on MRI intensity and texture. It also outperformed the classification results using the volume of putamen. 
% which represents the gold standard feature for the study of HD\xxx{ref to a paper or two using putamen volume in HD - whichever papers you're already citing in the intro}. 
In addition, our experiments indicate the complementary nature of putamen volume and tensor-based grading. 
%
%to preserve spatial coherency of patches and structural differences over different individuals. This results in better tracking of structural alterations occurring with Huntington’s disease.

% \xxx{if this is not anonymous, we also need an acknowledgement paragraph. Put my R01, and for the PREDICT grant.}

% \xxx{do refs count against your page limit? I forget. if so, i'm not sure that you need the first two refs here - you can use 3 every time you talk about hd}
\bibliographystyle{IEEEbib}
\bibliography{tensor-based_grading_ISBI2020}

\end{document}